\documentclass[%
twocolumn,
 fleqn,usenatbib,
 showkeys,
 floatfix,nolongbibliography,aps,
author-numerical%
]{revtex4-2}

\pdfoutput=1
\usepackage[T1]{fontenc}
\usepackage [latin1]{inputenc}
\DeclareRobustCommand{\VAN}[3]{#2}
\let\VANthebibliography\thebibliography
\def\thebibliography{\DeclareRobustCommand{\VAN}[3]{##3}\VANthebibliography}

\usepackage{newtxtext,newtxmath}
\usepackage{tikz,xcolor,hyperref}
\usepackage[normalem]{ulem}
\usepackage{graphicx,url,caption,subcaption}
\usepackage{dcolumn}
\usepackage{bm}

\draft 

\begin{document}


\definecolor{lime}{HTML}{A6CE39}
\DeclareRobustCommand{\orcidicon}{%
	\begin{tikzpicture}
	\draw[lime, fill=lime] (0,0) 
	circle [radius=0.16] 
	node[white] {{\fontfamily{qag}\selectfont \tiny ID}};
	\draw[white, fill=white] (-0.0625,0.095) 
	circle [radius=0.007];
	\end{tikzpicture}
	\hspace{-2mm}
}

\foreach \x in {A, ..., Z}{%
	\expandafter\xdef\csname orcid\x\endcsname{\noexpand\href{https://orcid.org/\csname orcidauthor\x\endcsname}{\noexpand\orcidicon}}
}

\newcommand{\orcidauthorA}{0000-0001-9180-4773}

\title[Interaction of electron beam with Harris CS]{Interaction of a spatially uniform electron beam with a rotational magnetic hole
in a form of a Harris current sheet}

\author{D. Tsiklauri\orcidA{}}
 \email{D.Tsiklauri@salford.ac.uk}
\affiliation{Joule Physics Laboratory,
School of Science, Engineering and Environment, 
University of Salford,
Manchester, M5 4WT, 
United Kingdom}

\date{\today}

\begin{abstract}
In this work we use
particle-in-cell (PIC) numerical simulations to study
interaction of a spatially uniform electron beam with 
a rotational magnetic hole in a form of a Harris current sheet.
We vary width of the Harris current sheet to investigate how this
affects the quasi-linear relaxation, i.e. plateau formation of
the bump-on-tail unstable electron beam.
We find that when width of the Harris current sheet
approaches and becomes smaller than the electron gyro-radius,
quasi-linear relaxation becomes hampered and a positive
slope in the electron velocity distribution function (VDF) persists.
We explain this by the effects of non-conservation of electron magnetic
moment, which, as recent works suggest, can maintain 
the positive slope of the VDF.
In part, this can explain why {{\it some}} electron beams 
{(the ones that interact with narrow magnetic holes with sharp
boundaries, represented in our study
by a Harris current sheet)}
in the solar wind
travel much longer distances than predicted by the quasi-linear
theory, at least in those cases when the electron beams slide along
the current sheets that are abundant when the different-speed solar
wind streams interact with each other.
\end{abstract}


\maketitle

\section{Introduction}

The two constituents of this work: (i) electron beams \citep{kavagh2012} and (ii) Harris current sheets \citep{thompson2005} are abundantly present in many space plasma situations.
There are many challenges in understanding complex
wave-particle interactions how electron beams interact with
solar wind structures such as density cavities \citep{thurgood2016}, 
magnetic holes \citep{karlsson2021,jiang2024}, 
current sheets \citep{asano2004} or how they affect the
heat flux carried into the heliosphere \citep{coburn2024}.
See \citet{tsurutani2023} for a recent review of the topic.

Solar wind observations at 1 AU show presence of
bump-on-tail unstable electron beams and associated Langmuir waves
generated by these beams \citep{lin1981}. 
{The more energetic electrons  naturally arrive
before the less energetic ones because of their higher speeds. 
These energetic electrons become spatially localized.
Due to these time-of-flight effects, 
the localized energetic electrons appear to "catch up" 
or even overtake less energetic electrons 
in a certain region, creating a localized "beam".
Thus, a bump-on-tail electron distribution function 
forms. This leads to a positive slope in the 
electron distribution function at a 
specific energy range (a bump), which is a 
source of free energy to drive Langmuir waves
\citep{muschietti1990}.}
Such velocity distribution function (VDF) is 
known to be unstable to the generation of Langmuir waves
in the regions of VDF where $\partial f / \partial v_\parallel > 0$. 
\citet{lin1981} computed
plasma wave growth 
from the distribution function and found that  it 
agrees with the observed onset of the Langmuir waves.
Also a qualitative agreement/coincidence was established 
between the variations in the Langmuir wave
observed amplitude  levels and in the regions of positive slope of VDF. 
The well-known problem is that
 evolution of the VDF, predicts much 
higher Langmuir wave amplitudes than what has been observed \citep{lin1981}.
This points to an existence of some kind of  a process
that impedes the quasi-linear relaxation of the electron beam to form
a plateau and it somehow maintains the positive slope of the VDF.
A theoretical explanation for such impediment of the interaction of the
beam and the 
background plasma, i.e. an effective de-resonation of
the electron beam from the generation of Langmuir waves was
offered before \citep{ryutov1969,breizman1970,nishikawa1976}.
In the framework of quasi-linear theory 
they have considered  how density fluctuations can suppress the
electron beam from the generation of Langmuir waves and found
the following two conditions
for this to happen:
(i) $\delta n/n_0 \geq 3 k^2 V_{\rm th,e}^2/\omega_{\rm pe}^2$,
or equivalently 
(ii) $\delta n/n_0 \geq 3 (V_{\rm th,e}/V_{\rm b})^2$.
{Here $\delta n$ and $n_0$ are 
number density fluctuation and background number density,
respectively, $k$ is Langmuir wavenumber, $V_{\rm th,e}=\left(
k T_e/ m_e \right)^{0.5}$ is electron thermal speed, $V_{\rm b}$
is the electron beam speed, $\omega_{\rm pe}=
\left(n_0 e^2/(m_e \varepsilon_0 \right)^{0.5}$ is electron
plasma frequency.}
Note that the second condition follows from the first one by simply
substituting $k \approx \omega_{\rm pe}/V_{\rm b}$, at the electron
beam and Langmuir wave resonance.
These conditions simply follow from Taylor expanding, (i.e. using 
$(1+\alpha)^n \approx 1+n \alpha$, for $\alpha \ll 1$),
the Langmuir wave dispersion relation that takes form:
$\omega_{\rm L}\approx 
\omega_{\rm pe}\left[1+\delta n/(2 n_0)\right]+3 k^2 V_{\rm th,e}^2/(2 \omega_{\rm pe})$,
and demanding that
$\omega_{\rm pe}\delta n/(2 n_0)\geq 
3 k^2 V_{\rm th,e}^2/(2 \omega_{\rm pe})$,
i.e. the relative number density ($\delta n/n_0$) exceeds the specified threshold.
\citet{thurgood2016} was the first work
that confirmed the findings of the quasi-linear theory by
Ryutov and co-workers 
\citep{ryutov1969,breizman1970,nishikawa1976} using fully kinetic particle-in-cell simulations.
\citet{thurgood2016}
found that when the background plasma density is homogeneous, 
Langmuir waves are resonantly generated with subsequent
quasi-linear relaxation
causing dynamic increase of the wavenumbers. 
In the homogeneous background plasma density 
case electrons are not accelerated. 
However, in the inhomogeneous plasma case,
when the density is spatially varying, and temporally static density
perturbations (density cavities) are added, then the electron acceleration
is seen. In  inhomogeneous case \citet{thurgood2016}
also found the generation of
backwards-propagating Langmuir waves, that are shown  
to be produced due to the
{\it refraction} of the wave packets from the edges of density cavities.
When the depth of the density cavities is above the above-mentioned
criteria ($\delta n/n_0 \geq 3 k^2 V_{\rm th,e}^2/\omega_{\rm pe}^2$),
properties of the generated 
Langmuir waves markedly differ from the usual Langmuir wave 
dispersion relation. Thus, density cavities were shown to efficiently
de-resonate electron beams from the Langmuir waves.

The bump-on-tail electron beam 
quasi-linear relaxation time scale 
can be calculated using Langmuir wave growth rate 
$\gamma_L=\pi ({\omega_{\rm pe}}/{n})v^2 {\partial f}/{\partial v_\parallel}$
\citep{melrose1980,McClements1986}. Simple arguments presented in
\citet{McClements1986} yield 
\begin{equation}
\gamma_L=\pi {\omega_{\rm pe}} \frac{n_{\rm b}}{n_0}\frac{v_{\rm b}}{\Delta v},
\label{eq1}
\end{equation}
where $n_{\rm b}/{n_0}$ is the ratio of electron beam to background
plasma number densities and ${v_{\rm b}}/{\Delta v}$ is the ratio
of electron beam speed to the effective width of the plateau 
(cf. Fig. 1 from \citet{McClements1986}). Eq.(\ref{eq1}) coincides
with the expression given in \citet{breizman1970}.
In most space plasma applications
${v_{\rm b}}/{\Delta v}\approx 1$, thus 
quasi-linear relaxation time, $T_{\rm QL}$, 
{which is defined as inverse Langmuir wave
growth rate, $1/\gamma_L$,
(in the units of 
inverse electron plasma frequency) can be estimated as
\begin{equation}
T_{\rm QL}\omega_{\rm pe}=
\frac{\omega_{\rm pe}}{\gamma_L}=\frac{1}{\pi}\frac{n_0}{n_{\rm b}}.
\label{eq2}
\end{equation}
}
For example, the solar type III radio bursts are produced by
very low density, $n_{\rm b} \approx (10^{-4}-10^{-6}) n_0$,
electron beams which, according to Eq.(\ref{eq2}) 
predict very long, quasi-linear relaxation time 
$T_{\rm QL}\omega_{\rm pe}=3\times (10^{3}-10^{5})$.
The conclusion reached by \citet{lin1981} {\it is} that,
observationally, electron beam quasi-linear relaxation time
{\it is} much longer than theoretical value predicted by Eq.(\ref{eq2}).

In addition to density fluctuations or density cavities in the
solar wind, there are similar structures also observed, {\it 
not} in density, but in the absolute value of magnetic field i.e.
$\sqrt{B^2}$ or equally in the square of 
magnetic field ${B^2}$. These structures are called {\it magnetic holes}.
Magnetic holes have been observed in the solar wind for many
decades starting from \citet{turner1977}.
\citet{turner1977} established the existence of
the two types of magnetic holes: 
(i) with a clear rotation of the magnetic field
{i.e. when the magnetic 
field one (or more) component(s) (and not the entire $B$!)
change(s) sign from a negative value
on one side of the hole to the positive one on the other side.}
These are so called rotational magnetic holes , and 
(ii) without such rotation, i.e. ones with a simple reduction in 
one or more components of the magnetic field. These are
so called linear magnetic holes. 
It should be noted that to our knowledge
{\it no publication exists which realizes
the similarity} of the rotational magnetic hole with
with the Harris \citep{harris1962} current sheet. 
Indeed similarly is very profound see e.g.
Fig. 3 from \citep{karlsson2021}
where in panel (d), red curve showing the magnetic field varies 
exactly as
in Harris current sheet.
Note that in addition to the classical
\citet{harris1962} equilibrium solution,
more advanced, kinetic-scale exact stationary solutions of
hybrid Vlasov-Maxwell equilibria for sheared plasmas with in-plane 
and out-of-plane magnetic field, without \citep{malara2018} and with
with inhomogeneous temperature \citep{malara2022}, also exist.
In this work, we exploit this similarity
because in the numerical simulations of
collisionless magnetic reconnection,
Harris current sheet has been very well studied,
literally being a 'work horse' for decades
due to the ease of its numerical implementation and stability,
see \citet{jan2015} as a representative example.
As far as the formation mechanisms for the two types of
magnetic holes are concerned, for the linear ones
the main possible mechanism is the mirror mode instability\citep{ahmadi2017},
while for the rotational ones there is no clear idea how
they form other than suggestions of a possible
magnetic reconnection at the current sheets with
the magnetic field rotation \citep{turner1977,zhang2008}.
The recent developments in the magnetic hole research
include kinetic-scale events 
with observations of the structure at a sub-proton scale \citep{gershman2016}. 
It was found that electrons with gyro-radii exceeding 
the thermal gyro-radius but smaller than 
thickness of the current layer can carry a current to
cause a 10-20\% reduction in magnetic field 
absolute value \citep{gershman2016}.

Based on  a semi-graphical framework
using the equations of quasi-linear theory to 
describe electron-driven instabilities in
the solar wind \citep{verscharen2022},
recent works suggest that magnetic holes
that have sharp spatial gradients or have deep
reduction in magnetic field  absolute value
can maintain 
the positive slope of the VDF both for the case of
production of Langmuir waves \citep{liu2025} and whistler 
waves \citep{jiang2025}.
\citet{liu2025} explains this by the effects of 
non-conservation of electron magnetic
moment which can maintain the positive slope of the VDF
in the velocity range just under the electron beam speed 
{{(See for details
The results section below for more details)}}.

In the above context, the main stated aim of this paper is to investigate the
interaction of a spatially uniform electron beam with 
a rotational magnetic hole in a form of a Harris current sheet,
using PIC numerical simulations. 

\section{The model}

We use fully kinetic, explicit, electromagnetic,
2D, particle-in-cell (PIC) code called EPOCH \citep{arber2015}.
The physical parameters in our numerical
simulation are set to be commensurate to
solar wind conditions.
In particular, we set the relevant {scales} of both
plasma number density as
$n_0=10^{6}$ particles per m$^{-3}$,
background magnetic field, that is along y-axis, as
$B_0=5.7$ nT.
They are both {(as well as gas pressure) 
varying across the Harris current sheet
as
\begin{equation}
n_0(x)=n_0/\cosh^2(x/\delta), \;\;\;
p_0(x)=p_0/\cosh^2(x/\delta),
\label{eq3}
\end{equation}}
\begin{equation}
B_{\rm 0 y}(x)=B_0\tanh(x/\delta),
\label{eq4}
\end{equation}
where $\delta$ is the current sheet width-parameter,
sometimes called half-width.
The relevant thermal pressure scale is
set as $p_0=B_0^2/(2 \mu_0)$, which ensures
that the plasma beta in the simulation is unity ($\beta=1$),
as it is expected for the solar wind plasmas.

Background temperature is spatially uniform 
{and is assumed to be the same
for
all plasma species in our PIC simulation. It
is set using the ideal gas law}, $p_0=n k_{\rm B} T_0$, i.e.
$T_0=p_0/k_{\rm B}/n_0$.
In effect, enforcing $\beta=1$ condition, and setting
$n_0$ and $B_0$ sets the temperature as 
$T_0=80.7$ eV or $T_0=9.4\times 10^5$ K,
which is also similar to the typical solar wind conditions.
The above equations ensure that the total pressure, i.e.
the sum of thermal and magnetic pressures
is indeed constant. 
{It is simple to verify that in the classical
Harris current sheet
\begin{equation}
\frac{B^2}{2 \mu_0}+p=\frac{B_0^2}{2 \mu_0}\tanh^2\left(\frac{x}{\delta}\right)
+ p_0{\rm sech}^2\left(\frac{x}{\delta}\right)=2 p_0=const,
\label{eq5}
\end{equation}
if $p_0=B_0^2/(2 \mu_0)$, which is indeed the case here
and the force balance in the current sheet holds
$\vec J \times \vec B = \vec \nabla p$ i.e.
\begin{equation}
\frac{B_0^2}{\delta \mu_0}
{\rm sech}^2\left(\frac{x}{\delta}\right)
\tanh\left(\frac{x}{\delta}\right)=\frac{d}{dx}
p_0{\rm sech}^2\left(\frac{x}{\delta}\right),
\label{eq6}
\end{equation}
as in \citep{harris1962}.}
The simulation box length is set as
$x_{\rm max}=n_x r_{\rm D}$, $y_{\rm max}=n_y r_D$
where $n_x=n_y=400$ is number of grids in x- and y-directions
and $r_{\rm D}=V_{\rm th,e}/\omega_{\rm pe}$ is the Debye radius.
The simulation domain span is set as
$-x_{\rm max}/2 \leq x \leq x_{\rm max}/2$ and
$-y_{\rm max}/2 \leq y \leq y_{\rm max}/2$.
Which means the grid cell length is set to the Debye radius
to ensure the numerical stability and to properly resolve
the relevant plasma kinetic effects.
Such domain size has the following properties:
$x_{\rm max}/(c/\omega_{\rm pe})=       5.03$ and 
$x_{\rm max}/r_{\rm L} =      7.11$, i.e.
the simulation box size is 5 times larger than electron inertial
length $c/\omega_{\rm pe}$ and 7 times larger than electron
Larmor radius (gyro-radius) $r_{\rm L}=V_{\rm th,e}/\omega_{\rm ce}$,
where $\omega_{\rm ce}=e B_0/m_e$ is the electron cyclotron frequency.
The chosen system size has a good reason:
as shown by \citet{liu2025} 
the adiabatic invariance of the electron 
magnetic moment $\mu=\gamma m_e V_\perp^2/(2B_0)$
is fulfilled when electron beam passes through a magnetic hole
that satisfies the following two conditions
(i) $r_{\rm L}/R_{\rm MH} \lessapprox 1$, where $R_{\rm MH}$ is the width (radius) of the
magnetic hole, i.e. Larmor radius of an electron is smaller than
that of the magnetic hole; and 
(ii) $\tau \omega_{\rm ce} \gtrapprox 1$, i.e. 
electrons must complete several gyrations while crossing
the magnetic hole over time $\tau$, 
{where $\tau$ is the magnetic hole crossing time.}
As shown by \citet{liu2025} observations of Solar Orbiter
corroborate that electrons beams in steep and deep magnetic
holes can violate conservation of $\mu$.
These conditions also imply that on contrary when
magnetic holes have sharp spatial gradients or have deep
reduction in magnetic field  absolute value,
these can re-shape the VDF to fulfill
the condition for the bump-on-tail instability
in the region of velocities just  under the beam speed to
have $\partial f / \partial v_\parallel > 0$.
Hence in cases when $r_{\rm L}/R_{\rm MH}>1$
$\mu$ is not conserved, which results
in the formation of the positive slope in the VDF.
Thus, our approach is to vary $\delta/ x_{\rm max}= 0.05, 0.1, 
0.15, 0.25, 0.5, 1.0$
such that this covers the range  $r_{\rm L}/R_{\rm MH}=1.60-0.16$.
Note that the value of 0.16 for this ratio is quoted for 
$\delta/ x_{\rm max}=0.5$ because for $\delta/ x_{\rm max}=1.0$
full width at half maximum (FWHM) for the number density
given by Eq.(\ref{eq3}), which we use as the width of the magnetic hole, 
$R_{\rm MH}$,
exceeds the simulation box size, hence FWHM estimation is not possible.
Commensurately, in Table \ref{t1},
for this reason, the Run 6, in column 
$R_{\rm MH}/x_{\rm max}$ just indicates $>1$ and
column $r_{\rm L}/R_{\rm MH}$ has no entry, just a bar.
For $\delta/ x_{\rm max}=0.05$ case, 
the method how we calculate $r_{\rm L}/R_{\rm MH}=
r_{\rm L}/[x(217)-x(182)]=1.60$ 
is that we find the
array indexes for which the number density drops by a factor of two, i.e.
$n(217)/n_0=n(182)/n_0=0.5$.
Here, for this value of $\delta/ x_{\rm max}=0.05$,
$R_{\rm MH}$ is equal to $[x(217)-x(182)]$.
Note that for these indexes also 
$[B_{\rm 0 y}(217)/B_0]^2=[B_{\rm 0 y}(182)/B_0]^2=0.5$.
For the $r_{\rm L}/R_{\rm MH}=1.60$ and generally 
$r_{\rm L}/R_{\rm MH} \gtrapprox 1$ we expect to see non-conservation of
electron magnetic moment $\mu$ and the formation/persistence of a 
positive slope in the VDF near and below the velocities of the electron beam.

We start PIC simulation with 4 plasma species, where
the first 
2 species are electrons and protons which maintain the Harris current sheet
with properties given by Eqs(\ref{eq3}-\ref{eq4}), which form
a background plasma.
For electrons we add momentum drift in z-direction as
$p_{\rm 0z}=- m_e B_0/(\delta \mu_0 e n_0) $ as required for
maintaining the usual Harris current sheet equilibrium.
Background ions have no such drift is current is maintained
by electrons only. Empirically this was found the produce
the most stable Harris current sheet that exists without
losing its shape for 1000s of plasma periods 
$T_{\rm pe}=2\pi/\omega_{\rm pe}$. 
The second additional  2 species we inject are 
beams of electrons and protons, which are (initially) spatially
uniform, have the number densities $n_{\rm b,e}=n_{\rm b,p}=0.005 n_0$,
and which have the following drift momenta
$p_{\rm 0y,e}=p_0$ and $p_{\rm 0y,p}=p_0 m_r$
for electrons and protons
where $p_0=m_e c \sqrt(\gamma^2-1)$. Here the chosen
$\gamma=1.0+3000 e/(m_e c^2)$ value sets the 
Lorentz $\gamma$ corresponding to the 3 keV electron beam.
This is a typical value for the solar wind electron beams.
Here $m_r=m_p/m_e=1836.153$ is the proton to electron mass ratio
used in our PIC simulations.
We sample electron distribution function
with 30000 grid points in the range $-2 p_0 \leq p \leq 2 p_0$,
which is more than enough to resolve its dynamic evolution.
As explained in \citet{tsiklauri2024},
it is important to ensure that initially a  zero net 
current is enforced. This avoids onset of spurious plasma
oscillations with electron plasma frequency $\omega_{\rm pe}$.
{In practice this means that
when the beam electrons are given initial momentum of $p_{y,e}=p_0$,
commensurately, the beam ions (protons) are given 
momentum $p_{y,i}=p_0 m_r$.
Note that when setting $J_z$-current that is needed to maintain
the Harris current sheet, the background electrons are
given momentum $p_{z,{\rm Harris}}=-(B_0/(\delta \mu_0 e n_0))m_e$, while
the background ions are stationary, i.e. electrons carry
all of the $J_z$-current}.
The choice of $n_{\rm b,e}=0.005 n_0$
is motivated by the two factors: (i) 
as explained in the Introduction section,
observationally electron beams have very
small number densities in the range
 $n_{\rm b} \approx (10^{-4}-10^{-6}) n_0$,
which, according to Eq.(\ref{eq2}) 
predicts a very long quasi-linear relaxation time of
$T_{\rm QL}\omega_{\rm pe}=3\times (10^{3}-10^{5})$.
Computationally such long simulation times are rather challenging.
Thus many authors chose unrealistically large ratios, such as
$n_{\rm b,e}/n_0 \geq 0.05 $.
However, as shown by \citet{thurgood2015} too large values of 
beam-to-background plasma number density result in
signification deviations from Langmuir dispersion 
relation characteristics of the initial wave modes
which generally have detrimental effect on 
satisfying frequency matching requirements 
(wave beat conditions) in the three-wave interaction processes.
In fact, \citet{thurgood2015} were the first to verify 
the arguments made from the theoretical perspective by 
\citet{cairns1989}. For recent developments see also \citet{bacchini2024}.
In this light, our choice of 
$n_{\rm b,e}=0.005 n_0$ is a reasonable compromise.
This sets the quasi-linear relaxation time for our PIC simulations
as  $T_{\rm QL}=1/0.005/\pi\approx64/\omega_{\rm pe}$.
This is, of course, should be regarded as an approximate value
given the number of the simplifications assumed in the theory. 

The boundary conditions (BC) used are as follows:
In x-direction: 
(i) electric and magnetic fields have imposed
open  BC. { When applied to EM fields, open BC imply
the waves out-flowing characteristics propagate through the
boundary, 
(ii) while particles obey periodic BC,
which means that particles reaching one edge of the
domain are wrapped round to the opposite boundary. So their total
number is strictly conserved.
}
In y-direction: the EM fields and particle BC are both periodic.
{{Open BC mean when applied to fields, the waves 
out-flowing characteristics propagate through the
boundary. Particles are transmitted through the boundary and removed from the system.
Periodic BC means fields and/or particles reaching one edge of the
domain are wrapped round to the opposite boundary.}}
Such BC ensure that: (i) particles are not lost from the simulation
domain (i.e. the total number of particles is conserved in the simulation), 
as they are periodic and 
(ii) at the same time
they allow for $B_{\rm 0 y}(x)$
variation across the Harris current sheet according to
Eq.(\ref{eq4}). Note that one cannot use periodic BC
for the EM fields for the Harris current sheet.

We use 100 particles per cell for each species, so 
in total,  we have 
$4 \times 100\times 400 \times 400 = 6.4\times10^7$
particles. In cases of long numerical runs with end
simulation time of $2500/\omega_{\rm pe}$,
a numerical run takes circa 
4 days on 32 processor cores on 
 Intel(R) Xeon(R) CPU E5-2630, 2.20GHz Linux computing node.
We assured the numerical convergence of the
presented results. 
In particular we checked twice finer grid and twice larger number of
particles yields numerical results which look the same at 
plotting accuracy. The length is normalized on Debye radius $r_{\rm D}$.
Note in the case of narrowest Harris current sheet with
$\delta/ x_{\rm max}=0.05$, it is resolved with 120 grid points.

A thought needs to be given to electric field
normalization. $E_0$ is set to
$E_0=0.25 \omega_{\rm pe} m_e c / e=24.04$ V m$^{-1}$.
When dealing with {\it EM waves} injected in plasma,
according to \citet{esarey2009}
a relevant electric field scale  is
so-called wave breaking electric field.
One usually defines a parameter $a=q_e E_0/(m_e \omega_0 c)$,
where $\omega_0$ is the frequency of the EM wave.
It is said that when
 $a\geq 1$, the electron quiver motion is relativistic and the
EM-plasma interaction is nonlinear. This leads
to wave breaking, i.e. its over-turning due to the nonlinearity).
For $a\leq 1$ the EM wave stays linear. 
In our case in the simulation we generate 
{\it Langmuir, electrostatic waves}, which have frequency $\omega_{\rm pe}$.
So we set $\omega_0=\omega_{\rm pe}$ therefore for 
our above value of $E_0$ 
$a$ is equal to 0.25, meaning that the results are in the 
linear regime.

In this work we present  6 numerical simulation cases, 
as specified in Table \ref{t1}.

\begin{table}
\captionsetup{justification=raggedright,
singlelinecheck=false}
\caption{Table of numerical runs considered.}
\centering
\begin{tabular}{lcccc} 
\hline
Run &  $\delta/ x_{\rm max}$ & $R_{\rm MH}/x_{\rm max}$ &$r_{\rm L}/R_{\rm MH}$&$t_{\rm end}\omega_{\rm pe}$ \\
\hline

1  & 0.05 &0.09&1.60 & 2500 \\
2  & 0.10 &0.17&0.81 & 2000 \\
3  & 0.15 &0.26&0.53 & 1750 \\
4  & 0.25 &0.44&0.32 & 1000 \\
5  & 0.50 &0.88&0.16 & 500  \\
6  & 1.00 &$>1$&---  & 500  \\
\hline
\end{tabular}
\label{t1}
\end{table}

\section{The results}

\begin{figure*}[!htb]
\captionsetup{justification=raggedright,
singlelinecheck=false}
\centerline{\includegraphics[width=0.95\textwidth]{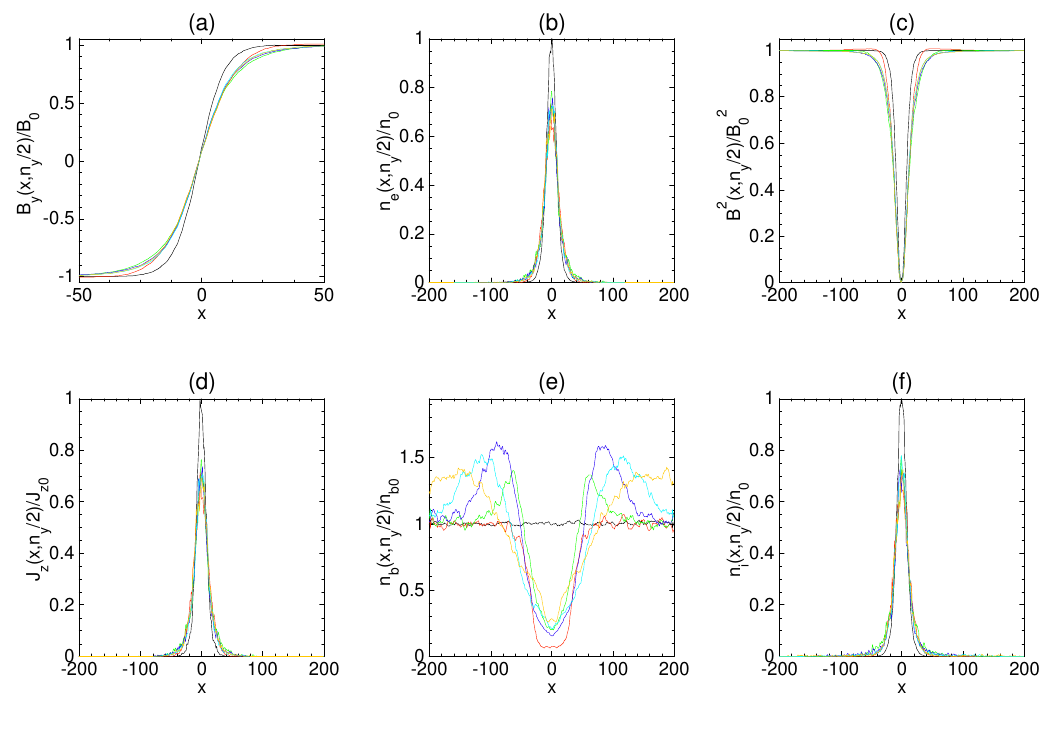}}
\caption{Time evolution of different physical quantity
profiles across the middle ($y=y_{\rm max}/2$) of the Harris current sheet:
(a) plot of  $B_y(x,y=y_{\rm max}/2,t)/B_0$ (please note
this panel is {\it zoomed-in} the x-range),
(b) background electron number density $n_e(x,y=y_{\rm max}/2,t)/n_0$,
(c) the total magnetic field squared 
$\sum_{i=x,y,z}[B_i^2(x,y=y_{\rm max}/2,t)]/B_0^2$, 
which shows the depth of the magnetic hole,
(d) out-of-plane current $J_z(x,y=y_{\rm max}/2,t)/J_{z0}$,
(e) electron beam number density $n_b(x,y=y_{\rm max}/2,t)/n_0$,
(f) background ion number density $n_i(x,y=y_{\rm max}/2,t)/n_0$.
The data is for Run 1, the narrowest current sheet $\delta/ x_{\rm max}=0.05$. 
See table \ref{t1} for details.
Black, Red, Green, Blue, Cyan, and Gold lines 
correspond to time instances of $t=0,0.2,0.4,0.6,0.8,1.0~t_{\rm END}$.}
\label{fig1}
\end{figure*}

\begin{figure*}[!htb]
\captionsetup{justification=raggedright,
singlelinecheck=false}
\centerline{\includegraphics[width=0.95\textwidth]{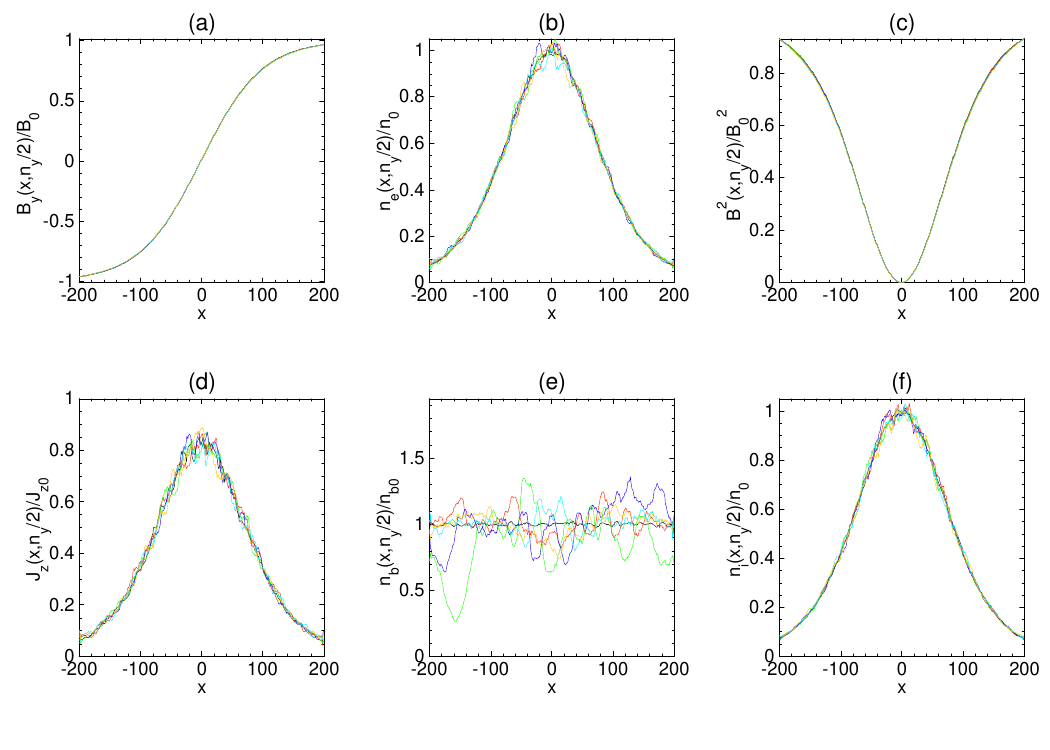}}
\caption{The same as in Fig.(\ref{fig1}), but for the Run 5,
a wide current sheet with $\delta/ x_{\rm max}=0.5$.}
\label{fig2}
\end{figure*}

Before we consider the subject matter, first we wish to
ascertain the stability of the Harris current sheet throughout
the numerical simulation. For this purpose, we 
produce Fig.(\ref{fig1}) where we present
the time evolution of different physical quantity
profiles across the middle ($y=y_{\rm max}/2$) of the Harris current sheet:
(a) plot of  $B_y(x,y=y_{\rm max}/2,t)/B_0$,
which is the Harris current sheet-associated
background magnetic field,
(b) the background electron number density $n_e(x,y=y_{\rm max}/2,t)/n_0$,
(c) the total magnetic field squared 
$[B_x^2(x,y=y_{\rm max}/2,t)+B_y^2(x,y=y_{\rm max}/2,t)+
B_z^2(x,y=y_{\rm max}/2,t)]/B_0^2$, with the purpose of
 showing  the depth of the magnetic hole,
(d) out-of-plane current $J_z(x,y=y_{\rm max}/2,t)/J_{z0}$,
(e) electron beam number density $n_b(x,y=y_{\rm max}/2,t)/n_0$,
(f) background ion number density $n_i(x,y=y_{\rm max}/2,t)/n_0$.
The data is for Run 1, the narrowest 
current sheet $\delta/ x_{\rm max}=0.05$. 
See table \ref{t1} for details.
Black, Red, Green, Blue, Cyan, and Gold lines 
correspond to time instances of $t=0,0.2,0.4,0.6,0.8,1.0~t_{\rm END}$.
Note the EPOCH does not output $J_z$ current at $t=0$. Thus,
in panel (d) for in Fig.(\ref{fig1}) and Fig.(\ref{fig2})
the black line is at the first snapshot, which is very close the start
of the simulation ($t=0.01t_{\rm END}$) but not precisely $t=0$.
In panel (a) we see that the background magnetic filed
is evolving rather slowly by $t_{\rm END}=2500 \omega_{\rm pe}$,
with its profile across the current sheet becoming less steep.
This is understood by a slow diffusion of the rather steep
current sheet. This serves as the proof that our background Harris
current sheet equilibrium is quite stable.
The panels (b), (d) and (f) demonstrate that  
the background electron number density $n_e(x,y=y_{\rm max}/2,t)/n_0$,
out-of-plane current $J_z(x,y=y_{\rm max}/2,t)/J_{z0}$, and
background ion number density $n_i(x,y=y_{\rm max}/2,t)/n_0$
have their maxima slowly reduced, as the simulation progresses.
Panel (c) shows that the magnetic hole width is slowly increasing
with time, which is consistent with panel (a) where 
 $B_y(x,y=y_{\rm max}/2,t)/B_0$ becomes less steep.
A particularly interesting panel is (e) where
we observe that initially electron beam is uniform across
the current sheet (Black line), then a {\it deep density cavity forms}
(Red line), which subsequently forms two maxima (that can be called
'wings' - Blue line). Subsequently as time progresses,
Green, Cyan and Gold lines indicate that the 'wings' spatially move
away from the center of the current sheet and its depth becomes
depleted. We speculate that this behaviour can be explained by
the pitch angle scattering \citet{liu2025} who
 explains this by the effects of 
non-conservation of electron magnetic
moment which can maintain the positive slope of the VDF
in the velocity range just under the electron beam speed.
Note that for this case 
$r_{\rm L}/R_{\rm MH}=1.6>1$ which means that the condition for
non-conservation of electron magnetic
moment {\it is fulfilled}.

{We would like to remark 
that in Run 1 the magnetic field evolves quite slowly and this serves as 
a proof that the background Harris sheet is indeed stable. 
One could surmise that
time evolution of $B_y(x)$ could be partially due to the presence of the electron-proton beam. However, this is not the case because of the two reasons
(i) the beam-to-background number density ratio is too small
$n_{\rm b,e}/ n_0=n_{\rm b,p}/ n_0=0.005 \ll 1$ to have any sizable effect;
(ii) we performed numerical runs without the beam, when
initial distribution function is in a form of a plateau,
i.e. without the beam but and have 
parameters corresponding to Run 1,
with a hope to see that non-conservation of electron magnetic
moment would create a positive slope in the simulation.
Such numerical runs (not shown here) did not show any positive slope 
formation i.e. the plateau remained flat. Which proves that the 
physical system's dynamics with or without the beam is not just additive
and involves some complex wave-particle interaction. 
Further additional analysis is needed,
which can be subject of a future work.
}

Next in Fig.(\ref{fig2}) we present the stability analysis of a 
wide current sheet with $\delta/ x_{\rm max}=0.5$.
We gather from this figure that in this case the time 
evolution of the current sheet is even slower, i.e.
in almost all panels the different color lines
overlap to the plotting accuracy. Again, very interesting panel
is panel (e) which demonstrates that 
when the condition for 
{non-}conservation of electron magnetic
moment is {\it not} satisfied
$r_{\rm L}/R_{\rm MH}=0.16\ll 1$,
there is {\it no density cavity formation}
in the electron beam, which demonstrates 
no effects related to the non-conservation of electron magnetic moment
 are taking place.

\begin{figure*}[!htb]
\captionsetup{justification=raggedright,
singlelinecheck=false}
\centerline{\includegraphics[width=0.95\textwidth]{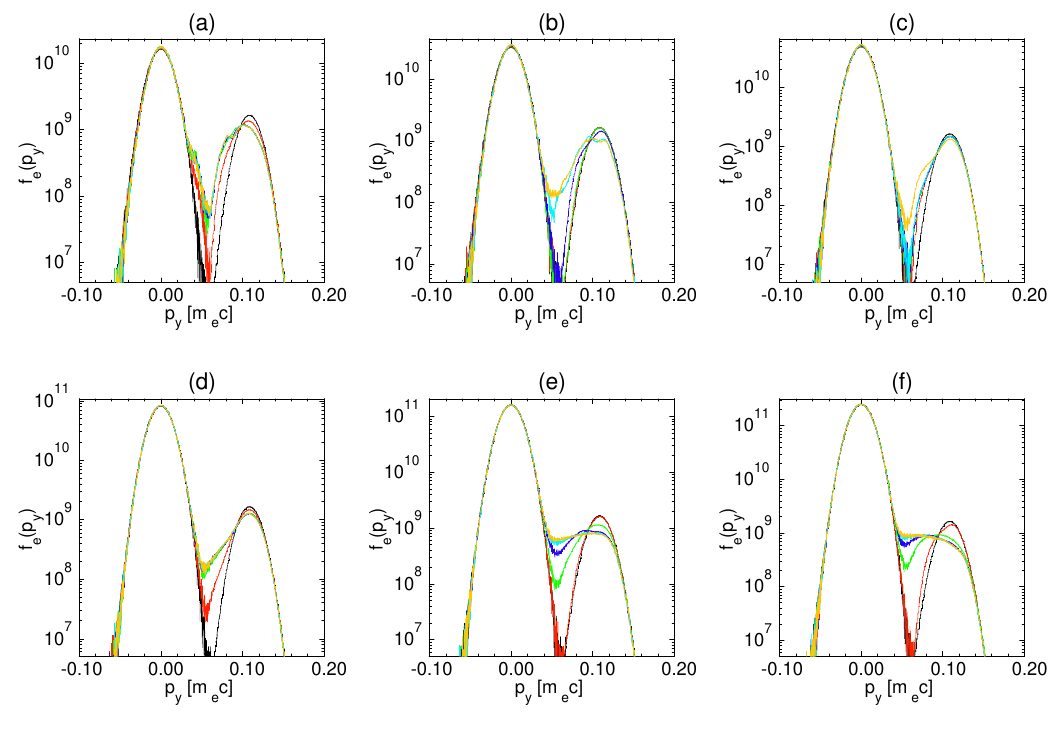}}
\caption{Time evolution of the sum of background and beam VDF
 parallel to the magnetic field component,
i.e. $f_e(p_\parallel)\equiv f_e(p_y)$.
(a) is for Run1, 
(b) is for Run2, 
(c) is for Run3,
(d) is for Run4,
(e) is for Run5,
(f) is for Run6. 
Here, again Black, Red, Green, Blue, Cyan, and Gold lines 
correspond to time instances of $t=0,0.2,0.4,0.6,0.8,1.0~t_{\rm END}$.
Note that $t_{\rm END}$ is different for each run.}
\label{fig3}
\end{figure*}

After discussing the stability analysis of the two extreme cases of
current sheets, we now move on to present the main subject matter results
by plotting 
the time evolution of the sum of background and beam VDF
 parallel to the magnetic field component,
i.e. $f_e(p_\parallel)\equiv f_e(p_y)$ for the different
numerical Runs 1-6 in Fig.(\ref{fig3}).
{The VDF is calculated using EPOCH code's built-in
capability to integrate over all spatial coordinates, i.e.
we produce and investigate VDF for the entire simulation domain.}
Panels (a)-(d) have a common feature in that as the time progresses
VDF parallel to the magnetic field component always
maintains a positive slope. This is especially evident in panel
(a) for which the condition for
non-conservation of electron magnetic
moment is fulfilled, i.e. $r_{\rm L}/R_{\rm MH}=1.6>1$.
We gather that Red curve starts to show a beginning of the
quasi-linear relaxation, but as the time passes
Green, Blue, Cyan and Gold color lines overlap as
if the time evolution stops and VDF has always a positive slope
$\partial f / \partial p_\parallel \equiv \partial f / \partial p_y > 0$.
Note that the quasi-linear relaxation time for our PIC simulations
as  $T_{\rm QL}=1/0.005/\pi\approx64/\omega_{\rm pe} \ll
 t_{\rm END}=2500\omega_{\rm pe}$ for
Run 1. The same ($T_{\rm QL} \ll
 t_{\rm END}$) is true for all other numerical runs.
In  panels (b), (c) and (d) we see similar
behaviour as in panel (a), but less drastic. For these
three cases, while $r_{\rm L}/R_{\rm MH}$ has the following values
$0.81,0.53,0.32$, which are not strictly speaking satisfying
$r_{\rm L}/R_{\rm MH} \gtrapprox 1$ i.e. the condition for
non-conservation of electron magnetic
moment, but the values are not far from unity.
The drastic difference  in the behaviour appears in
panels (e) and (f) where Gold color line
shows a clear plateau formation, which is an indication 
of the end of the quasi-linear relaxation process.
That is why we actually stopped the simulation in these two cases at 
$t_{\rm END}\omega_{\rm pe}=500$. Note that in panel (f) the 
Gold color line even shows a small negative slope, this could be
due to the finite difference effects present in the numerical simulation
causing a numerical dissipation (damping).

\begin{figure*}[!htb]
\captionsetup{justification=raggedright,
singlelinecheck=false}
\centerline{\includegraphics[width=0.95\textwidth]{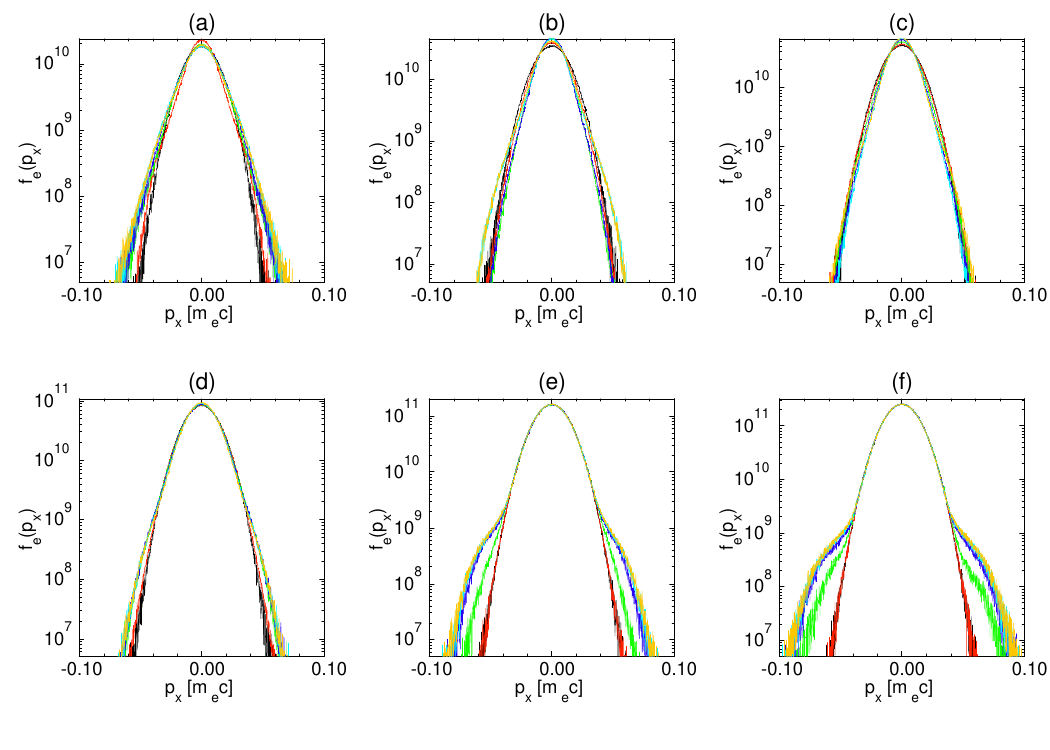}}
\caption{The same as in Fig.(\ref{fig3}), but for 
the background and beam VDF
 {\it perpendicular} to the magnetic field component
$f_e(p_\perp)\equiv f_e(p_x)$.}
\label{fig4}
\end{figure*}

In Fig.(\ref{fig4}) we show
the time evolution of the sum of background and beam VDF
 perpendicular to the magnetic field component.
An interesting common pattern that can be deduced is
that in panels (a)-(d) we see small widening of the
VDF which is an indication of the increase
in the perpendicular temperature, $T_\perp$.
The cause of this heating is the 
double layer-type
perpendicular electric
field $E_\perp \equiv E_x$, which as will be discussed
later in  Fig.(\ref{fig8}) attains  
values of
$E_x\simeq 5 \times 10^{-3}$ that are comparable to $E_\parallel \equiv E_y$.
An interesting observation can be also made about
panels (e) and (f). We gather that Red, Greed, Blue and Gold
curves show progressive formation of {\it electron beams}
in the perpendicular direction.
The cause of these beams is the 
wave-like perpendicular electric
field $E_\perp \equiv E_x$, which as will be discussed
later in  Fig.(\ref{fig9}).
Similar electron beams were also seen (see Fig. 5(a) 
from \citep{tsiklauri2012}) in their PIC simulation,
but formed in the direction parallel to the background
magnetic field. PIC simulations  of \citet{tsiklauri2012}
prove (see their Figs. (6)--(8)) that when 
such beams are plotted on the log-log 
plots of electron {\it energy} spectra they appear as the 
'knee', frequently seen in the solar flare observations, and that they
can be interpreted is the Landau damping of 
inertial Alfven waves due to the wave-particle interactions.
The physical nature of these wave seen in $E_\perp \equiv E_x$, 
in  Fig.(\ref{fig9}) needs to be investigated further.

\begin{figure*}[!htb]
\captionsetup{justification=raggedright,
singlelinecheck=false}
\centerline{\includegraphics[width=0.95\textwidth]{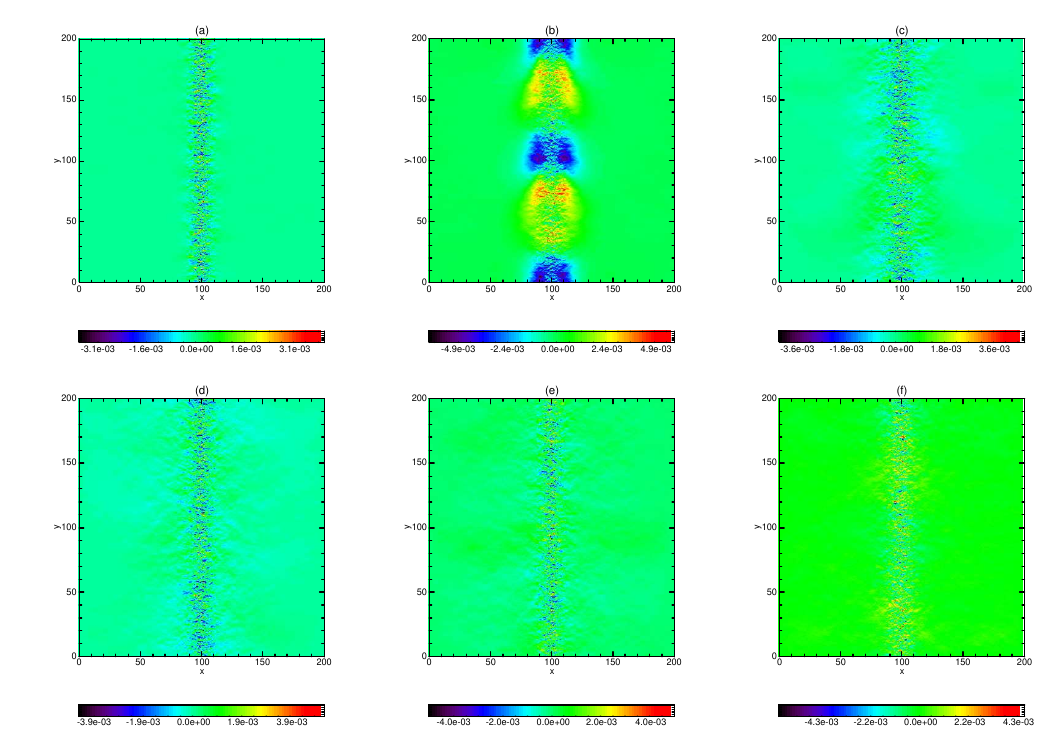}}
\caption{Time evolution of electric field component
parallel to the background magnetic field, i.e.
electric field associated with Langmuir waves, $E_y$.
The data is for Run 1, the narrowest current sheet $\delta/ x_{\rm max}=0.05$. 
Panels (a), (b), (c), (d), (e), (f)
correspond to time instances of 
$t=0.01,0.2,0.4,0.6,0.8,1.0~t_{\rm END}$.
The electric field is normalized to 
$E_0=0.25 \omega_{\rm pe} m_e c /q_e$. See text for details.}
\label{fig5}
\end{figure*}

Since it is known that Langmuir waves are generated
by the bump-on-tail unstable VDF, next,
in Fig.(\ref{fig5}) we explore the
time evolution of electric field component
parallel to the background magnetic field, $E_y$.
{In principle, it would be better
 to show evidence for Langmuir waves in the 
 frequency or wave vector domain. For example, 
 it would be illustrative to show the 
 Fourier-transformed dispersion plots using 
 the simulated electric-field data.
Producing such plot requires to store the data
snapshots every $0.2\omega_{\rm pe}$, i.e. five points per wave period.
At the moment the data is stored at much coarser $1.0\omega_{\rm pe}$
time resolution. Thus computational cost overweights the benefit of
producing the Fourier-transformed dispersion plots. }
Here the data is for Run 1, 
the narrowest current sheet $\delta/ x_{\rm max}=0.05$.
We make the following observations from panels (a)-(f) that:\\
(i) the electric field is confined to the very narrow
current sheet, essentially where the plasma exists.\\
(ii) the electric field  has very small, noise-like, spatial
scales, such that no clear wavelength can be deduced. This is probably
because of the complicated processes of wave-particle
interactions in the regime  of
non-conservation of electron magnetic
moment.\\
(iii) panel (b) stands out by showing relatively
large spatial-scale structures (yellow and blue blobs),
which we attribute to the possible electron vortexes
that can be identified at the same time in
Fig.(\ref{fig7}) also in panel (b).
{These vortexes may be formed by 
{Kelvin-Helmholtz (KHI)} the two-stream instability (TSI) 
as the substantial shear between Harris current sheet 
background electrons and the injected
electron beams exists.
It should be noted that vortexes form
both in the cases of
Kelvin-Helmholtz (KHI) and TSI.
The main difference is that KHI 
occurs at the interface of two distinct fluids 
due to a velocity shear across that interface
(including a discontinuous jump between two distinct velocities), 
causing them to roll into vortices, 
while the TSI occurs when two inter-penetrating 
streams of particles in a plasma have different drift velocities, 
leading to instabilities driven by their relative motion. 
KHI creates a mixing layer with 
visible rolls and waves, whereas TSI involves one stream 
of particles flowing through another, such as electrons drifting through ions in a plasma. 
For these reasons we conjecture that TSI is at play in our case.
We also remark that prolific vortex formation and their suppression
by the background magnetic was found for the {\it electron scale} 
KHI (ESKHI) \citet{tsiklauri2024}, which is very different from its hydrodynamic
and magnetohydrodynamic analogs. 
After commenting on the differences between KHI and TSI the
similarities should be also stated:
note that the {\it analytical} growth rate of ESKHI, Eq.(30)
(see also corresponding Figure 3) from \citet{Alves_2014} is
identical to Eq.(11) (see also corresponding Figure 2) from \citet{2508.14362}
 replacing discontinuous velocity 
jump between two distinct velocities $\pm v_0$ 
by the electron beam speed $v_{0b}$.}
{We gather from
Figs. \ref{fig5} and \ref{fig7} a chain of vortexes appear. 
These structures seem to have a dominant wavelength 
corresponding to approximately the half domain size.
The scale of vortices in TSI
 is determined by the wavenumber of the unstable modes, 
 which is influenced by the relative drift speed, 
 and the physical properties of the streaming fluids. 
 For instance, a larger relative drift speed 
 typically leads to smaller vortices (larger wavenumbers).
TSI is described by a spectrum of 
unstable modes, each with a specific wavenumber. 
The smallest wavelengths (largest wavenumbers) 
or longest wavelengths (smallest wavenumbers) 
that grow fastest determine the scale of the resulting vortices, 
as shown in the above referenced figures in
\citet{Alves_2014,2508.14362}.
A separate analysis is needed to settle this issue, which is beyond
the scope of this work.
}

Note that because in EPOCH electric field is strictly zero 
at $t=0$, in Fig.(\ref{fig5}), Fig.(\ref{fig8}) and 
Fig.(\ref{fig9}) in the respective panels (a) 
we use $t=0.01t_{\rm END}$ 
at the first snapshot, 
which is very close the start
of the simulation, 
but not precisely $t=0$.

\begin{figure}
\captionsetup{justification=raggedright,
singlelinecheck=false}
\begin{center}
  \includegraphics[width=\columnwidth]{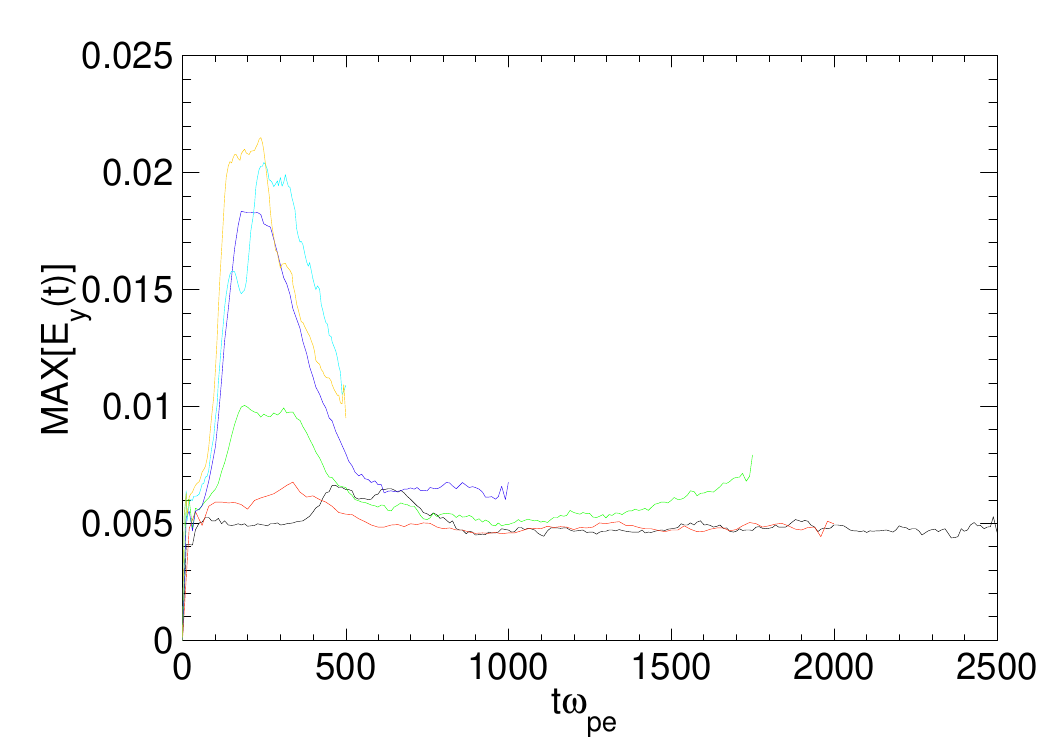}
\end{center}
\caption{Time evolution of the maximum of 
electric field component
parallel to the background magnetic field, i.e.
electric field associated with Langmuir waves, $\max[E_y(x,y,t)]$.
Black, Red, Green, Blue, Cyan, and Gold lines 
correspond to Run 1, Run 2, Run 3, Run 4, Run 5, Run 6, respectively. }
\label{fig6}
\end{figure}

In Fig.(\ref{fig6}) we study how 
the maximum of 
electric field component
parallel to the background magnetic field, i.e.
electric field associated with Langmuir waves, $\max[E_y(x,y,t)]$
changes with time.
Black, Red, and Green curves attain rather small maximal
values. These are the cases when the 
non-conservation of electron magnetic
moment  maintains the positive slope of the VDF.
Thus, Langmuir waves do not have a chance to grow
in amplitude markedly because the quasi-linear
relaxation is hampered by the wave particle
interaction.
On contrary, 
Cyan and Gold color curves show that
 when the electron beam forms a plateau,
see for reference panels (e) and (f) from Fig.(\ref{fig3}),
the Langmuir waves attain significant amplitudes.
This is because quasi-linear
relaxation has an opportunity to grow the waves without an
impediment from the effects of non-conservation of electron magnetic
moment.
{We thus ought to 
conclude that when $\mu$ is not conserved the 
instability itself is suppressed. Of course, in the absence of 
instability there is no quasi-linear relaxation, 
but one should not think that the lack of quasi-linear 
relaxation causes a lack of Langmuir wave growth seen in 
Fig.(\ref{fig6}).}

\begin{figure*}[!htb]
\captionsetup{justification=raggedright,
singlelinecheck=false}
\centerline{\includegraphics[width=0.95\textwidth]{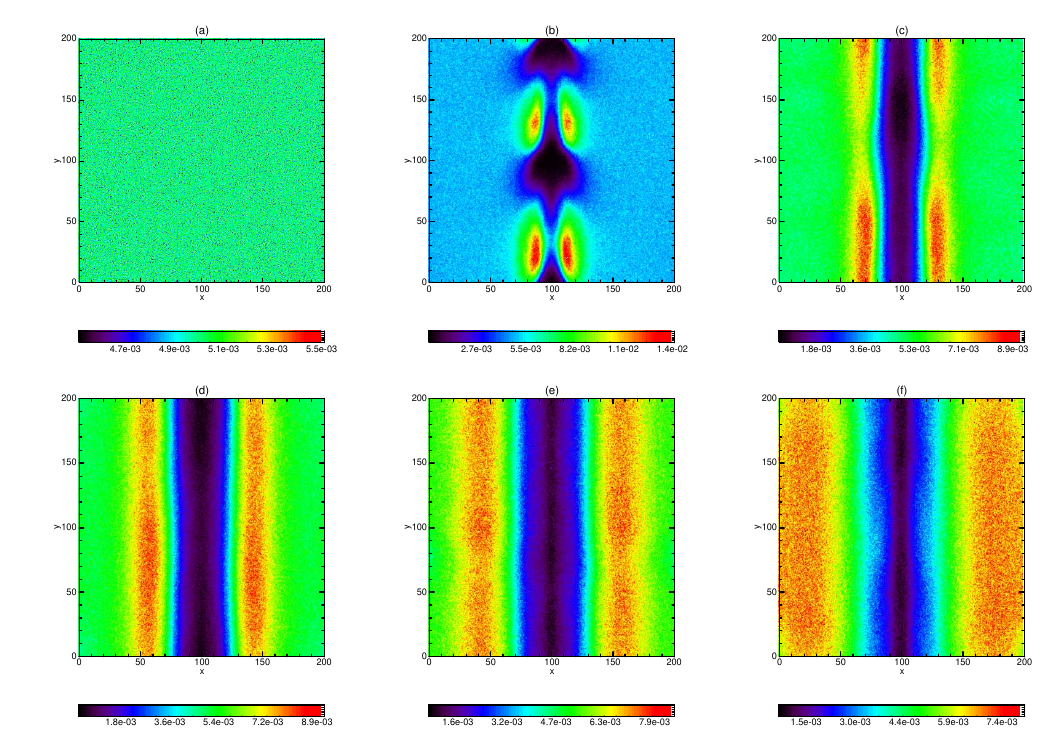}}
\caption{Time evolution of electron beam number
density $n_b(x,y,t)$.
Note that this is {\it excluding background number density, just the beam
number density is plotted.}
The data is for Run 1, the narrowest current sheet $\delta/ x_{\rm max}=0.05$. 
Panels (a), (b), (c), (d), (e), (f)
correspond to time instances of 
$t=0,0.2,0.4,0.6,0.8,1.0~t_{\rm END}$.}
\label{fig7}
\end{figure*}

In Fig.(\ref{fig7}) we present the
time evolution of electron beam number
density $n_b(x,y,t)$.
We see from panel (a) that the electron beam evolution starts
from a spatially uniform state at $t=0$.
In panel (b) we find  
structures (yellow and blue blobs),
which can be understood as electron vortexes
that are formed by { TSI} instability, because
there is a sufficient shear between Harris current sheet 
background electrons and the injected
electron beams. Moreover, this conjecture
is supported by the fact that when we plot
$J_y$  (not shown here for brevity)
significant current y-component exists {\it right in the middle}
of the current sheet, i.e. right in-between the vortexes.
The difficulty with EPOCH is that it does not readily
output plasma bulk velocity which can visualize
the existence of a shear flow. So, one should
use current  $J_y=-e n_e V_{e,y}$ instead to infer
 bulk velocity $V_{e,y}$ that clearly show the shear.
 In panels (c), (e), and (f) we subsequently 
 see the formation of a cavity (the blue channel
 in the middle), flanked by the over-dense
wings (yellow lanes on the both sides).
These over-dense wings (of flanks) widen as the time
progresses, this behaviour  can be also seen in panel (e)
of  Fig.(\ref{fig1}).

\begin{figure*}[!htb]
\captionsetup{justification=raggedright,
singlelinecheck=false}
\centerline{\includegraphics[width=0.95\textwidth]{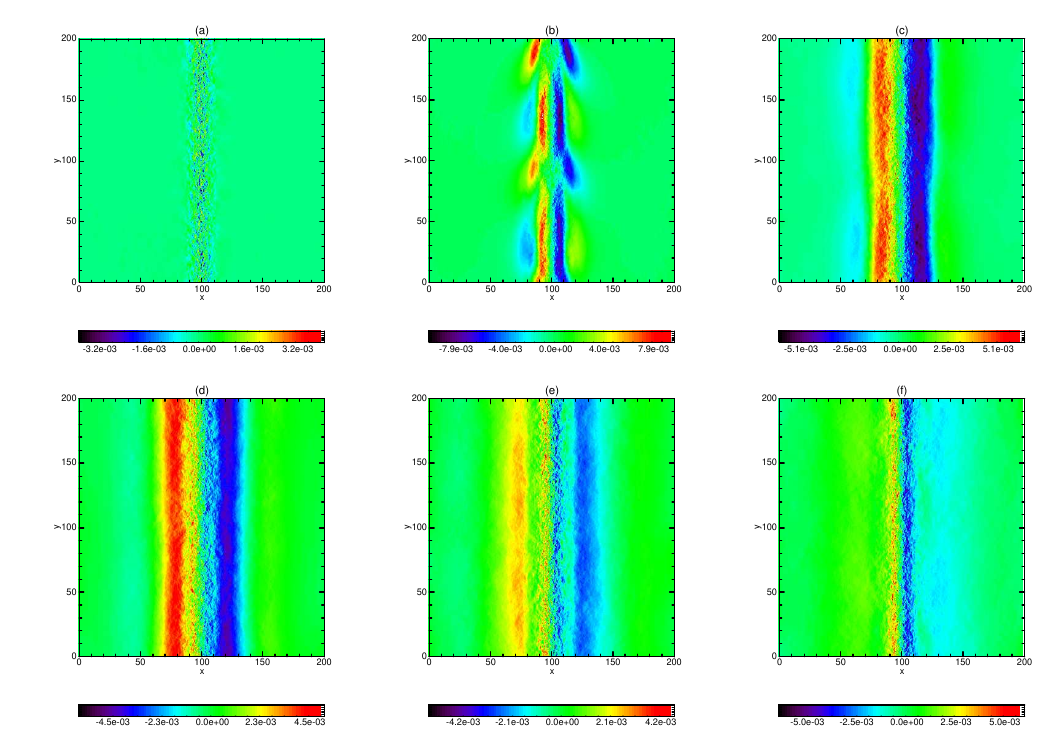}}
\caption{Time evolution of electric field component
perpendicular to the background magnetic field, $E_x$.
Panels (a), (b), (c), (d), (e), (f)
correspond to time instances of 
$t=0.01,0.2,0.4,0.6,0.8,1.0~t_{\rm END}$.
The data is for Run 1, the narrowest current 
sheet $\delta/ x_{\rm max}=0.05$.}
\label{fig8}
\end{figure*}

In Fig.(\ref{fig8}) we visualize the 
time evolution of electric field component
perpendicular to the background magnetic field, 
$E_x$. Note that 
here the data is for Run 1, the narrowest current 
sheet $\delta/ x_{\rm max}=0.05$.
This has the purpose to 
understand what cases the 
 the increase
in the perpendicular temperature, $T_\perp$
in panels (a), (b), (c) and (d) seen 
in Fig.(\ref{fig4}). 
We see that as the time progresses
double layer-type
perpendicular electric
field $E_\perp \equiv E_x$ forms.
It is interesting to note in panel (b)
that this double layer is disrupted by the 
electron vortexes. The largest amplitude
of $E_\perp =5 \times 10^{-3}$ 
is attained in panel (d), and it subsequently
fades away.
We surmise that the heating and electron beam formation
seen in Fig.(\ref{fig4}) is caused by this perpendicular
electric field.

\begin{figure*}[!htb]
\captionsetup{justification=raggedright,
singlelinecheck=false}
\centerline{\includegraphics[width=0.95\textwidth]{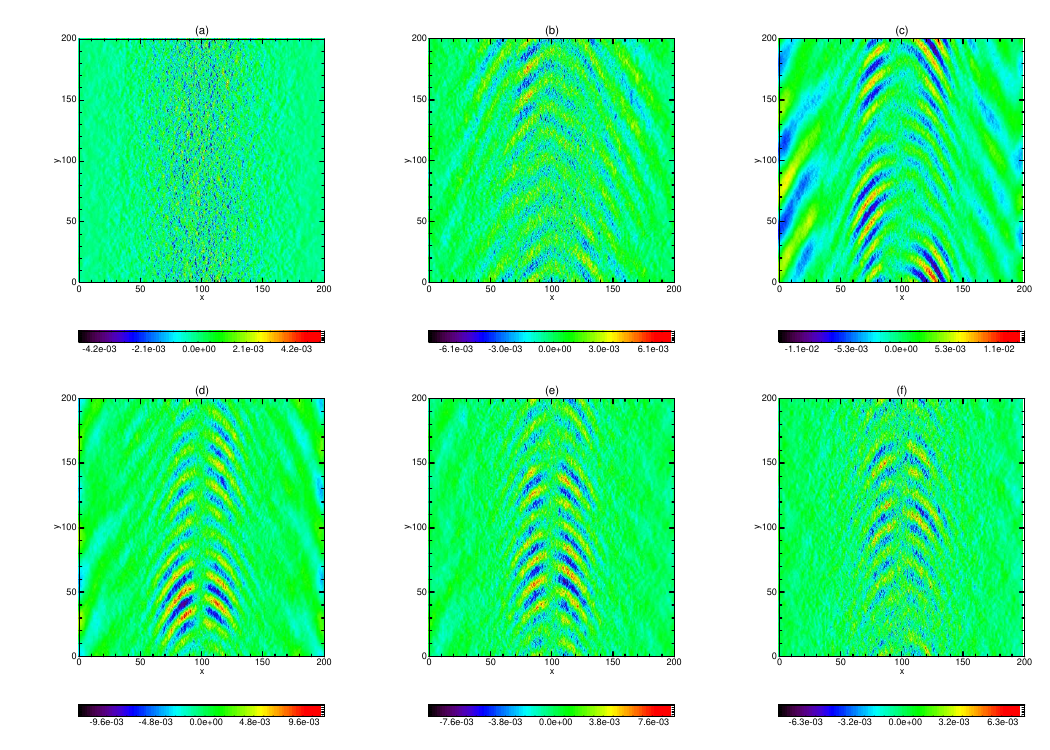}}
\caption{Time evolution of electric field component
perpendicular  to the background magnetic field, i.e.
$E_x(x,y,t)$.
Panels (a), (b), (c), (d), (e), (f)
correspond to time instances of 
$t=0.01,0.2,0.4,0.6,0.8,1.0~t_{\rm END}$.
The data is for Run 5, a wide current sheet $\delta/ x_{\rm max}=0.5$.}
\label{fig9}
\end{figure*}

In Fig.(\ref{fig9}) we show
 the time evolution  of 
electric field component
$E_x(x,y,t)$.
Note that 
here the data is for Run 5, a wide current 
sheet $\delta/ x_{\rm max}=0.5$.
Previously we have  have seen the
formation of electron beams seen 
in panels (e) and (f) in Fig.(\ref{fig4}).
The cause of these beams can be understood 
{by the presence of} 
wave-like (or fishbone-like) perpendicular electric
field $E_\perp \equiv E_x$,
shown in Fig.(\ref{fig9}) in panels (b), (c) and (d).
The electric field subsequent decay is seen
in panels (e) and (f).

\begin{figure}
\captionsetup{justification=raggedright,
singlelinecheck=false}
\begin{center}
  \includegraphics[width=\columnwidth]{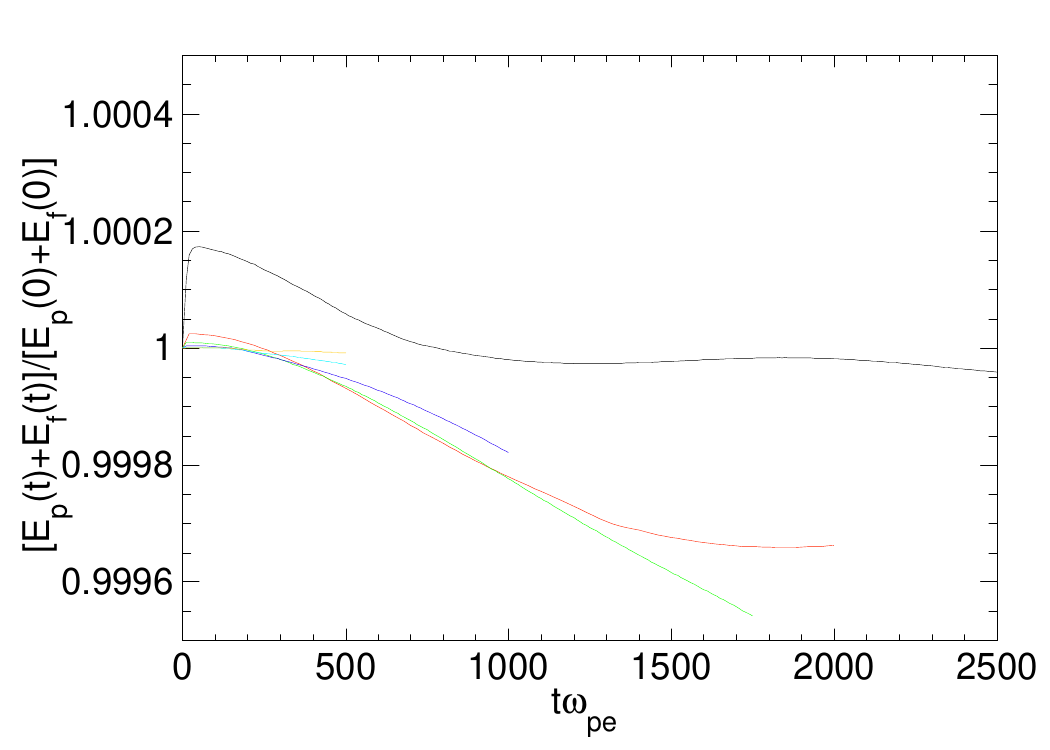}
\end{center}
\caption{Similar to Figure \ref{fig6} but now for 
    the total energy, i.e. the sum of all particle kinetic energy plus the
 electric and magnetic energies, normalized to the initial value at
 $t=0$.}
\label{fig10}
\end{figure}

To conclude, in Fig.(\ref{fig10}) 
we ascertain the total energy conservation in all our numerical
simulations. Here we plot
the sum of all particle kinetic energy plus the
 electric and magnetic energies,  normalized to the initial value at
 $t=0$.
It can be seen in the figure that the normalized 
total energy
variation is within the range of
$[0.9996-1.0002]$. This indicates a nearly perfect
total energy conservation, which means the numerical
resolution of the spatial grid and the number of particles
per cell is indeed adequate. The values
above unity are usually attributed to numerical
heating (errors), or indeed a real, physical 
instability, while the values 
of below unity are explained by the
 numerical diffusion caused by the finite differencing.
All of these effects are within a tolerable
margin, confirming the accuracy of our numerical simulations. 

\section{Summary and conclusions}

In this work we address a simple question,
not answered before: how the
variation of the 
width of the Harris current sheet 
affects the quasi-linear relaxation, that is 
the plateau formation of
the bump-on-tail unstable electron beam.

For this purpose we employed 
particle-in-cell (PIC) numerical simulations to study
interaction of a spatially uniform electron beam with 
the rotational magnetic hole in form 
of a Harris current sheet.
To our knowledge
 no previous work exists which notes/acknowledges
the similarity of the rotational magnetic hole with
with a Harris current sheet. 
Indeed similarity is very profound.
As can be seen in Fig. 3 from \citep{karlsson2021}
 in panel (d), red curve, that corresponds to the case of a
rotational magnetic hole, the magnetic field varies as
in a typical Harris current sheet.  
We considered 6 numerical runs see
 Table \ref{t1} for details, 
$\delta/ x_{\rm max}= 0.05, 0.1, 
0.15, 0.25, 0.5, 1.0$
such that this covers the range  
$r_{\rm L}/R_{\rm MH}=1.60-0.16$. 

By considering a narrow (Fig.(\ref{fig1})) and wide (Fig.(\ref{fig2})) 
Harris current sheets
we ascertained their stability  throughout
the numerical simulation. 
The narrow-wide division is whether
the non-conservation of electron magnetic
moment {\it is} (narrow case) or  {\it not} (wide case) taking place.
In the case of narrow current sheet we found that
initially uniform across
the current sheet electron beam forms a deep density cavity,
which subsequently forms two maxima, the 'wings', that later spatially move
away from the center of the current sheet and its depth becomes
depleted. We speculate that this behaviour can be explained 
by the effects of 
non-conservation of electron magnetic
moment that maintains the positive slope of the VDF,
in the velocity range just 
under the electron beam speed \citep{liu2025}.
In the case of wide current sheet 
electron beam does not form such a density cavity.

We find that in the narrow current sheet cases,
as the time progresses
VDF parallel to the magnetic field component always
maintains a positive slope (Fig.(\ref{fig3}) (a)-(d)).
This is because  when the width of the Harris current sheet
approaches and becomes smaller than the electron gyro-radius,
quasi-linear relaxation becomes hampered and the positive
slope in the electron velocity distribution function (VDF) persists.
In the wide current sheet cases 
the drastic difference in the behaviour is that
a clear plateau formation can be witnessed 
(Fig.(\ref{fig3}), (e), (f)), which is an indication 
of the end of the quasi-linear relaxation process for the wide current sheets.

We find that in the narrow current sheet cases,
a small increase in width of the
VDF which is an indication of the increase
in the perpendicular temperature, $T_\perp$, (Fig.(\ref{fig4}) (a)-(d)) and
the cause of this heating is the 
double layer-type perpendicular electric
field $E_\perp \equiv E_x$ (Fig.(\ref{fig8})).
In the wide current sheet cases
we see progressive formation of electron beams (Fig.(\ref{fig4}), (e), (f))
in the perpendicular direction and the cause of these beams is the 
wave-like perpendicular electric
field $E_\perp \equiv E_x$  (Fig.(\ref{fig9})).

We find that
in the case of narrow current sheet
electric field parallel (Fig.(\ref{fig5})(b)) and
perpendicular  (Fig.(\ref{fig8})(b)) to the background
magnetic field components,
as well as
electron beam number density (Fig.(\ref{fig7})(b)),
show 
the formation of  electron vortexes, in the beginning
of the numerical simulation.
These vortexes {are likely to be caused by the TSI} 
instability.

As discussed in the Introduction,
\citet{thurgood2016} provided the first robust evidence via
PIC simulation that
sufficient depth density cavities can
de-resonate electron beams from the Langmuir waves.
In this work we show that similar effect can be achieved by
rotational magnetic holes. This can potentially
explain why some electron beams in the solar wind
travel much longer distances than predicted by the quasi-linear
theory. Of course, our argument works {{\it only} in
those cases when electron beams slide along
Harris(-like) current sheets
which represent narrow magnetic holes with sharp
boundaries,} which are often encountered
 when the different speed solar
wind streams collide. 
{Thus, the magnetic-hole scenario considered 
in this manuscript cannot explain the frequent presence 
of electron beams generally seen in the solar wind. 
Indeed, it is commonly accepted that electron beams 
or strahl electrons widely exist in most of 
the solar wind without magnetic holes or any 
other types of current sheets. 
Their existence is more likely to be related 
to electron transport (such as e.g. time-of-flight 
\citep{muschietti1990} and other propagation effects) in the heliosphere.}

\section*{Acknowledgements}
{The author would like to thank two anonymous referees
for their useful comments that improved this article and 
Dr. Earl E. Scime (Editor) and Dr. Alain Hilgers (Guest Editor) 
for organizing a useful and meaningful review process.}

\vskip 0.25cm
{\bf Data availability statement.}
The data and numerical codes 
that support the findings of this study are available
from the corresponding author upon reasonable request.



\nocite{*}
\bibliography{paper87}

\end{document}